\documentclass[%
 reprint,
superscriptaddress,
 amsmath,amssymb,
 aps,
]{revtex4-2}
\usepackage{graphicx}
\usepackage{dcolumn}
\usepackage{bm}

\begin{document}

\preprint{APS/123-QED}

\title{Phase Transition Points and Classical Probability}

\date{\today}

\author{Yonglong Ding}
\affiliation{School of Physics, Beijing Institute of Technology, Beijing 100081, China}
\affiliation{Beijing Computational Science Research Center, Beijing 100193, China}

\date{\today}

\begin{abstract}
 In order to gain a deeper understanding of complex systems and infer key information using minimal data, I classify all configurations based on classical probability, starting from the dimensions of energy and different categories of configurations. By utilizing the principle of maximum entropy, it is concluded that all possible configurations with the same energy have equal probabilities of occurrence. By using different representations of high and low energy, the emergence of a transition point has been inferred. Finally, I take the Ising model as an example and calculate the transition point of the thermodynamic phase transition, which is determined to be 2.25. This value is very close to the simulation value obtained by Monte Carlo method, but I have only consumed a small amount of computational resources in the process.

\end{abstract}

\maketitle


\section{\label{sec:level1}Introduction}

The study of phase transition points has always been an important topic of strongly correlated systems, and the algorithms proposed for strongly correlated systems have been applied to physics, chemistry and other fields, providing great insight into understanding the nature of these phenomena. In all these algorithms, the main ones that have attracted attention are Monte Carlo's algorithm and the algorithm of tensor networks. These two algorithms offer powerful tools for challenging strongly correlated problems in fields such as high-energy physics\cite{degrand2006lattice}, condensed matter physics\cite{ceperley1995path,foulkes2001quantum}, nuclear physics\cite{carlson2015quantum}, and chemistry\cite{hammond1994monte,needs2020variational}, and the algorithm itself has continued to develop in recent decades. However, some of the problems faced by these two algorithms have not been well solved.

As we use in physics simulations, any computational technique have their limitations and drawbacks. For example, Quantum Monte Carlo (QMC)\cite{sandvik2007evidence,melko2008scaling,jiang2008antiferromagnet} methods suffer from the “sign problem”, which means that the probability distribution being sampled is not positive-definite, leading to statistical noise and slow convergence. This issue becomes particularly severe when simulating interacting systems with fermions.\cite{mondaini2022quantum} The accuracy of tensor network algorithms decreases as the size of the system increases. This limitation is due to the exponential growth of the tensor dimensions, making it computationally infeasible to simulate large systems\cite{orus2014practical}.

Here, instead of solving these problems, we have made some attempts to develop new methods in order to supplement existing calculation methods. Our research subjects started with the Ising model. The Ising model exhibits a phase transition at a critical temperature, where the system undergoes a sudden change in its magnetic properties. Below the critical temperature, the spins are ordered and aligned in the same direction, producing a net magnetization. Above the critical temperature, the spins become disordered and randomly oriented, and the net magnetization approaches zero.

The principles of quantum mechanics allow for exploring a wide variety of possible states for microscopic particles. For Ising model, the spin of a particle can be in a state called “spin up” or “spin down,” which describes the orientation of the particle’s spin relative to a chosen axis. In quantum statistical mechanics, the distribution function is described by the Fermi-Dirac or Bose-Einstein distribution, which applies to interacting particles with integer or half-integer spin, respectively. So, could we calculate the probability of multiple particles appearing in various states directly? This way can be challenging because it requires knowing the exact positions and momenta of all the particles in the system, which is generally impossible. In addition, the probability of observing a particular configuration of particles can depend on the interactions between the particles and the system’s environment, making the problem even more complex. However, by using the fundamental laws of physics, we can categorize all possible states and directly calculate some of these special classes. By combining these calculations with our understanding of related physical laws, we can then solve for phase transition points. This approach allows us to keep computational requirements within an acceptable range and provides us with a powerful tool for studying complex systems and their behavior.
In the Ising model, we first classify all possible states based on the average spin of all particles. We then categorize the states further based on their energy levels. However, when dealing with a large number of particles, it becomes impractical to determine the number of configurations possible for each energy level and spin state directly.

To address this problem, we discovered that the number of configurations possible for each energy level and spin state is highest further away from the central point in a lower energy location, and lower for higher energy locations, where the number of configurations is highest closer to the central point. However, at very low energies, the maximum value point may not necessarily appear at the furthest point from the center, it may be very close. The transition between energy states is modeled by moving from high to low energy, and each transition corresponds to a change from one calculated situation to another.

As we move closer to the phase transition point, the energy gap between the states becomes smaller, resulting in an upper bound to the gap in the number of configurations between the states. This means that there exists an energy level at which the number of configuration changes from edge to center most slowly compared to all energy levels. We conjecture that this point is closely related to the phase change point.

\section{Thoery}
\begin{figure}[hptb]
\includegraphics[width=8.5cm]{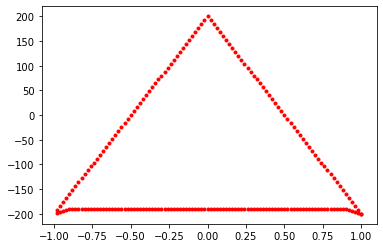}
\caption{ In the case of J=1 and n=10, the lowest energy configurations calculated for spin configurations of varying proportions are not parallel to the x-axis, but rather exhibit an initial increase with increasing x, followed by a convergence towards parallelism with the x-axis. When approaching a proportion of 1, the configurations initially exhibit a decrease and ultimately exhibit symmetry with the configurations on the left.(inside figure).}
\label{fig1}
\end{figure}

\begin{figure}[hptb]
\includegraphics[width=8.5cm]{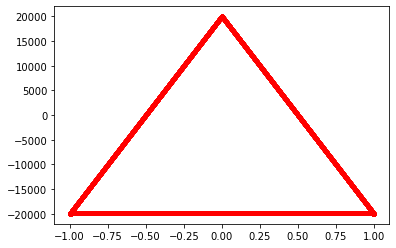}
\caption{ When J=1 and n=100, the lowest energy configurations calculated for spin configurations of varying proportions are relatively closer to parallelism with the x-axis compared to those in Figure 1.(inside figure) (inside figure). }
\label{fig2}
\end{figure}

\begin{figure}[hptb]
\includegraphics[width=8.5cm]{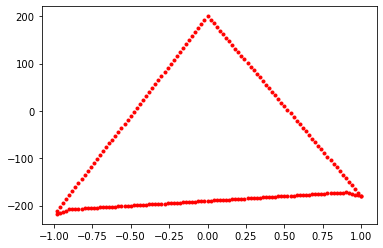}
\caption{ When J=1 and n=10, and an external magnetic field of B=0.2 is applied, there is a small upward lift on the right end compared to Figure 1.(inside figure). }
\label{fig3}
\end{figure}

\begin{figure}[hptb]
\includegraphics[width=8.4cm]{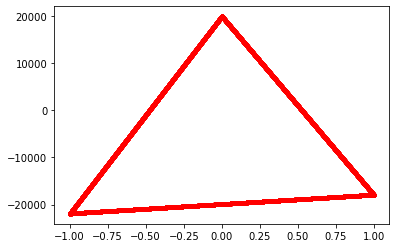}
\caption{When J=1 and n=10, and an external magnetic field of B=0.2 is applied, there is a small upward lift on the right end compared to Figure 2.}
\label{fig4}
\end{figure}

In the Ising model, individual particles are typically depicted as spins that can be in one of two states, whereas in Glass models, individual particles can possess a significantly greater number of states. However, observing the state of individual particles in a real system is considerably more challenging than observing the state of the system as a whole. To establish a relationship between the possible states of an individual particle and the possible states of the entire system, the approach taken here is to average the possible states of all particles directly. By utilizing the aforementioned method, we are able to determine the range of potential states for the entire system. When representing the spin direction as 1 and -1, the overall average value for both the Ising model and the spin glass model ranges from -1 to 1. Our algorithm initially categorizes the system based on the spin average, which can also be interpreted as a means of distinguishing based on upward magnetization strength, as illustrated in the aforementioned examples Fig.~\ref{fig1}-Fig.~\ref{fig4}.

 Periodic boundary conditions are particularly useful for Monte Carlo simulations of systems that are large enough to be affected by boundary effects, but for which it is computationally expensive to simulate the entire system. By assuming periodicity, the simulation can be run on a smaller “unit cell” of the system, which reduces the computational cost while still capturing the essential behavior of the larger system. We use the same periodic boundary conditions as Monte Carlo's algorithm, but only calculate the energy within the “unit cell”.

Given a Hamiltonian quantity, we can compute the energy of various configurations and determine the range of possible energy values. So, we subsequently categorize the states that were previously classified based on their energy magnitude. This provides the range of potential values for all possible states in both the spin average and energy dimensions.

Moving forward, let’s focus on the Ising model, beginning with the scenario of zero magnetic field strength.

\begin{equation}
H=-J\sum_{i,j}S_{i} S_{j},
\end{equation}
In the preceding discussion, we categorized all potential states based on the dimensions of spin average and energy. With J=1 and n=10, we are now able to generate a diagram.as show in Fig.~\ref{fig1} When the value of n is considerably large,
When J$>$0 and n is significantly large,as show in Fig.~\ref{fig2} the magnitude of energy is plotted on the vertical axis, and the degree of spin up and down is represented on the horizontal axis, where 1 and -1 denote spin up and down, respectively. The transverse axis is determined by calculating the sum of all spins and dividing the result by the number of spins. As depicted in the above figure, in the absence of an external field, the left and right sides are symmetric, which means the state with the lowest energy is when all spins are either up or down. For n=10, a line can be observed near the horizontal axis. The line, which is close to the horizontal axis and represents the minimum energy state, starts increasing slowly from x=-1 as the horizontal axis increases, approaches the straight line near x=1, and then gradually decreases. For large values of n, this line almost overlaps with the x-axis in the above figure.
When a magnetic field strength is present, the Hamiltonian changes to:
\begin{equation}
\begin{split}
H=-J\sum_{i,j}S_{i} S_{j}+B\sum_{k}S_{k};S_{k}=\pm1
\end{split}
\end{equation}
Given J=1 and n=10 with a magnetic field strength of B=0.2, we can plot,as show in Fig.~\ref{fig3} .
When n takes a relatively large value,as show in Fig.~\ref{fig4}.

Due to the applied magnetic field strength being greater than 0, the energy of the system is higher when all spins are facing up compared to that when they are all facing down. The probability of energy in various states will vary at different temperatures. At low temperatures, the Ising model will tend to have lower values on the vertical axis, while at higher temperatures it will tend towards higher values. In the above figure, bosons and fermions exhibit different characteristics.

We have classified all possible states that the model can produce, and in order to link them to the phase transition points, further analysis and calculations are required. Once the ratio of the mean spin up and the energy magnitude has been determined, corresponding to the values of the abscissa and ordinate in the above figure, further analysis and calculations are needed to establish their connection to the phase transition points. When the ratio of mean spin up to energy magnitude has been determined, which corresponds to the values of the abscissa and ordinate in the above figure, further analysis and calculations can be performed to understand the phase transition points. Points on the horizontal and vertical axes may have multiple solutions, which can be calculated using classical probability. However, these solutions are computationally intensive. To overcome this problem, we propose to calculate the values corresponding to only a small number of points, and then analyze the overall distribution using the trend of the values in different locations. A valid trend must incorporate the maximum entropy principle\cite{shao2023progress,gull1984maximum,silver1990maximum,gubernatis1991quantum,jarrell1996bayesian}.

The maximum entropy principle is a fundamental concept in statistical mechanics, which is the branch of physics that studies the behavior of large collections of particles. It states that, when we don’t have complete information about a system, the best way to make predictions about its behavior is to choose the probability distribution that has the highest entropy, subject to any constraints that we do know. We assume that all configurations with the same energy have an equal probability of appearing in a quantum system.
\section{ISING MODEL}
Taking the Ising model as an example, we aim to investigate the number of corresponding configurations at points located at different locations. To achieve this, we calculate the number of configurations for the three points of the triangle depicted above. The solutions for the left and right points are obviously unique, whereas the vertex has two possible solutions. Next, we consider two cases to explore more general laws. The first case involves selecting an energy near the bottom to calculate the corresponding configuration. In the second case, an energy near the vertex is selected to explore the distribution of states. Then, we demonstrate that the number of configurations generated decreases towards the center as the ratio of spin-up decreases, with the maximum value being found at or near the boundaries.

\begin{figure}[hptb]
\includegraphics[width=8.4cm]{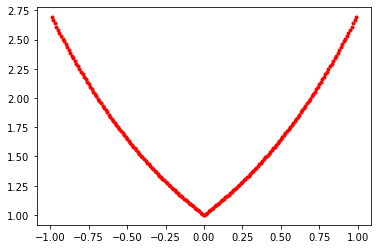}
\caption{ This figure is a schematic diagram illustrating the principle. When the energy is very low and close to the minimum energy point, the different types of configurations that can be generated for a given energy are mainly concentrated at the two ends. }
\label{fig5}
\end{figure}

\begin{figure}[hptb]
\includegraphics[width=8.5cm]{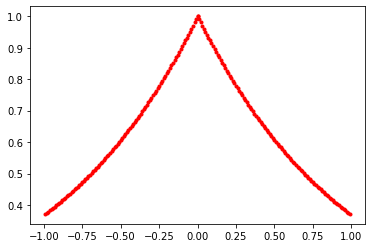}
\caption{ This figure is a schematic diagram illustrating the principle. When the energy is very high and close to the maximum energy point, the different types of configurations that can be generated for a given energy are mainly concentrated in the middle. }
\label{fig6}
\end{figure}

When given a specific energy, the leftmost point becomes the extreme point, and in this scenario, every lattice point with a spin-up orientation is encircled by points with a spin-down orientation. If you move slightly to the right, the number of lattice points with spin up increases by 1, and two cases arise: three particles connected together, or two pairs of two particles. We set the probability of m particles with upward spin surrounded by spins pointing downward as K. Additionally, we denote the probability of a scenario where m-1 particles have upward spin surrounded by downward spins as k1. In the aforementioned scenario, we only need to flip another lattice point with surrounding spins pointing upward to satisfy the conditions required to obtain K. In the scenario where there are m-2 particles with an upward spin orientation, to maintain energy conservation, we can only select to flip two lattice points around the m-2 particles with an upward spin orientation. In the case of very low energy as show in Fig.~\ref{fig5} , the number of possible configurations obtained in the former scenario is greater than that in the latter. We can obtain the scenario in the figure by mathematical induction.

And when the energy is relatively high we can get Fig.~\ref{fig6}.

When the energy takes its maximum value in the Ising model, the lattice points with upward spin orientation and those with downward spin orientation alternate. Each lattice point with upward spin is surrounded by those with downward spin, while each lattice point with downward spin is surrounded by those with upward spin. Flipping the lattice points with upward spin orientation in this configuration yields the same scenario as that in the case of very low energy. Moving from the middle to the left, the number of possible configurations obtained while maintaining the same energy decreases.

When the energy difference is extremely small, the distribution difference of configurations will also be minimal since flipping a spin induces an energy change less than 4. Thus, each configuration can be achieved by flipping the spin with energy differences of either 2 or 4. Therefore, when the energy magnitudes are relatively close, we can infer that the distribution of configurations is similar. We can infer the existence of a transition point between the two situations described above. Below this point, the configurations are mainly concentrated near the boundary, while above it, the configurations are mainly concentrated at the origin.

Based on the above inference, assuming that the critical point occurs at a specific energy level, the configuration exhibits higher symmetry at this level. Through the use of detailed balancing and other rules, we are able to calculate this point. Further, we believe that this point is very close to the phase transition point, and we can approximate the critical point by finding this particular point.

\section{CRITICAL POINT}
To compute all configurations for a given vertical axis $E_{i}$, we can calculate the distribution type of Figure 3 and Figure 4, which corresponds to $E_{i}$. The configurations presented are temperature-independent, and the total number of these configurations, denoted as $C_{i}$, can be counted. If we sum up the energy associated with each $C_{i}$, we can obtain all possible configurations, and the distribution of configurations corresponding to each energy level can be recorded as $C_{i}$ relative to the total. The magnetic field distribution for each energy is represented by mi. The calculation of the average energy can be expressed as follows:
\begin{equation}
	\begin{split}
		E=E_{1}C_{1}+E_{2}C_{2}+E_{3}C_{3}+...+E_{i}C_{i}+...
	\end{split}
\end{equation}

And since $C_{i}$ is a constant that does not change with temperature, we can derive a new formula for calculating energy. We can calculate the average value of different quantities using the following equation.
\begin{equation}
	\begin{split}
		M=E_{1}C_{1}M_{1}+E_{2}C_{2}M_{2}+E_{3}C_{3}M_{3}+...+E_{i}C_{i}M_{i}+...
	\end{split}
\end{equation}

\begin{figure}[hptb]
\includegraphics[width=8.5cm]{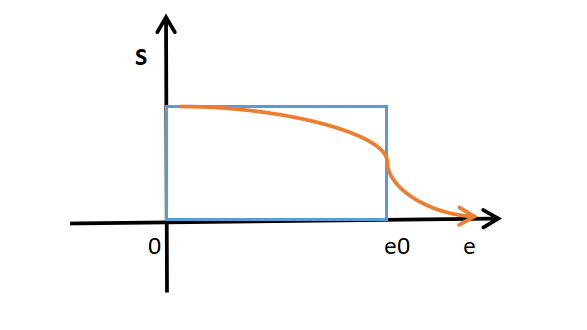}
\caption{ As the energy increases, the spins change from being mainly concentrated at the two ends to being concentrated in the middle, where there is a transition point e0 that can be approximated as the critical temperature of the phase transition. }
\label{fig7}
\end{figure}

The next problem we face is how to calculate phase transitions and other issues. Firstly, since the value of $C_{i}$ is determined, we can calculate it directly. The ratio of each energy configuration can be calculated by the partition function. When the temperature is fixed, the ratios of these configurations are also determined accordingly. For the Ising model with a scale of n, the value of $C_{i}$ can be calculated from the n x n matrix composed of 1 and -1 representing the Ising model. However, this method is computationally intensive.

Because I directly investigate the direct relationship between the phase transition point and the transition point between the two graphs found above.
Firstly, we derive M.
Because the magnetic field strength above the transition point is approximately 0, while below the transition point, the magnetic field strength has a larger value.

We consider the strength of the magnetic field, and it is evident that below the transition point, m decreases as E increases. However, the strength of the magnetic field has an image of values close to the edge. Above the transition point, the value of m decreases as E increases, but the magnetic field strength is close to the midpoint. Similarly, below the transition point, m also decreases with increasing E, but the magnetic field strength is close to the edge. At the exact transition point, the value of m is in the middle. It shows that the derivative of m with respect to E reaches its maximum value at the transformation point as show Fig.~\ref{fig7}.

The size of the area can be approximately obtained from the above figure as $E_{i}$*1, where $E_{i}$ corresponds to the energy at the transformation point. Therefore, the result obtained in the previous section is the point $E_{i}$. Therefore, the phase transition point is approximately equal to the energy transformation point.
\begin{figure}[hptb]
\includegraphics[width=8.5cm]{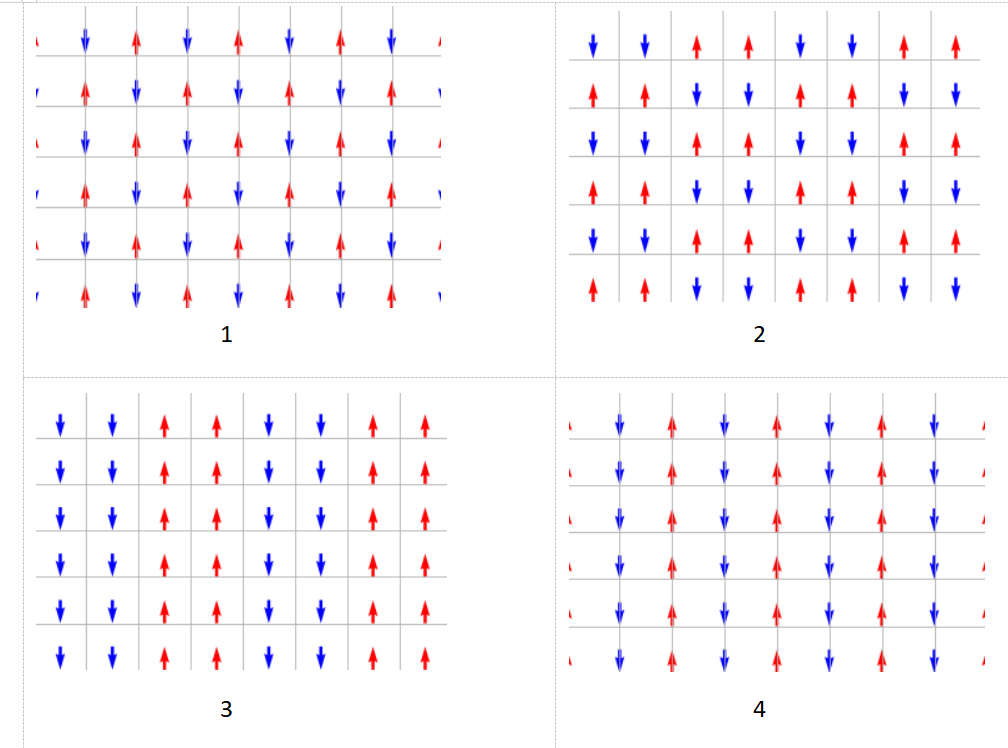}
\caption{ The four different types of spins are shown above, distributed according to different energies. Figure 1(m2) represents a spin with four neighboring spins different from itself, with an energy of 4. Figure 2(m1) represents a spin with three neighboring spins different from itself, with an energy of 2. Figure 3(m5) represents a spin with two neighboring spins different from itself, with an energy of 0. Figure 4(m3) represents a spin with one neighboring spin different from itself, with an energy of -2.}
\label{fig8}
\end{figure}

\begin{table}[b]
	\caption{\label{tab:table4}%
		   The following table represents the relationship between different configurations. When m1 is decreased by 4, m2 and m3 increase by 1 and 3, respectively. When m2 is decreased by 1, m4 increases by 1. When m3 is decreased by 4, m1 and m5 increase by 2, respectively. When m4 is decreased by 1, m2 increases by 1. When m5 is decreased by 4, m3 and m4 increase by 3 and 1, respectively.}
	\begin{ruledtabular}
		\begin{tabular}{ccddd}
			m1&m2&
			\multicolumn{1}{c}{\textrm{m3}}&
			\multicolumn{1}{c}{\textrm{m4}}&
			\multicolumn{1}{c}{\textrm{m5}}\\
			\hline
			-4&+1&\mbox{+3}&\mbox{}&\mbox{}\\
			{}& -1&{}& +1 & {} \\
			+2&{}&-4& {} & +2 \\
			{}&+1&{}& -1 & {} \\
			{}&{}&+3& +1 & -4 \\
		\end{tabular}
	\end{ruledtabular}
\end{table}
We are considering the configuration distribution of a certain energy, where the x-axis is near 0. In this situation, the probability of the spin distribution being up and down is almost equal. We want to obtain the derivative of this point. Flipping a spin from up to down results in an energy change of 4 if the flipped spin is surrounded by spins that are all pointing down. However, if there are only three neighboring spins pointing downwards, the energy change is 2. 
Near the zero point, we summarized the following five situations. The proportions of each situation are respectively marked as $m_{1},m_{2},m_{3},m_{4}$ and $m_{5}$ as show in Fig.~\ref{fig8}. 

If we flip the particles in the graph, we can obtain the relationships shown in the table 1

The table 1 shows the situations where flipping particles in the five configurations results in other particles, with the positive and negative signs indicating the increase and decrease of particles, respectively.
From the table 1, we can easily obtain a particularly stable situation where
\begin{equation}
	\begin{split}
		m_{1}+m_{5}=1.5m_{3}
	\end{split}
\end{equation}
and m2 and m4 does not exist$m_{2}=m_{4}=0$.
By considering the principle of detailed balance. As a result, we obtain
\begin{equation}
	\begin{split}
		m_{5}=e^{-2/T}m_{1},
		m_{4}=e^{-4/T}m_{2}
	\end{split}
\end{equation}
So by approximating temperature as the average energy of particles, we can obtain the equation 7.
\begin{equation}
	\begin{split}
		m_{1}+2m_{2}-2m_{4}-m_{5}=T-2
	\end{split}
\end{equation}
\begin{equation}
	\begin{split}
		m_{1}+m_{2}+m_{3}+m_{4}+m_{5}=1
	\end{split}
\end{equation}

As a result, we obtain
T=2.25

\section{Conclusion}
\label{conclusion}
Using classical probability, we can categorize all possible configurations. For the Ising model, for example, we can classify them based on the ratio of different particles that may appear, which is the proportion of spins pointing upward. Similarly, when the Heisenberg equation is given, we can obtain the energy of different configurations. In this way, from the perspectives of energy and configuration categories, we can classify all possible generated configurations. Furthermore, we also observed a gap between the graph of the Ising model and the x-axis. Based on this observation, we plotted the graph and considered two scenarios of different scales. We found that as the scale increases, the gap almost disappears. We further considered the case of an external magnetic field, in which case one of the angles of the triangular shape is slightly lifted.
Next, we calculated two scenarios where the energy is close to the highest and lowest possible values, and inferred that, under a given energy, the number of configurations that can be generated gradually decreases as the proportion of spins pointing downward decreases from 1 to 0.5 in the case of very low energy. However, for cases where the energy is very high, the number of configurations that can be generated continuously increases as the proportion of spins pointing downward decreases from 1 to 0.5.Because each configuration can be generated by flipping the nearby energy, we infer that the shape of the graph does not change dramatically as the energy increases. Based on this, we infer the existence of a transition point, which we associate with the phase transition point.
To determine this transition point, we assume that this transition occurs at a given energy. We take the Ising model as an example and solve it specifically using detailed balance, and finally draw a conclusion.

\nocite{*}

\bibliography{apssamp}

\end{document}